\documentclass[prl,aps,amssymb,showpacs,twocolumn]{revtex4}
\usepackage{amsmath}
\usepackage{amssymb}
\usepackage{amsthm}
\usepackage{amsfonts}
\usepackage{enumerate}
\usepackage{latexsym}
\usepackage{graphicx}

\newcommand{\beq}{\begin{equation}}
\newcommand{\eneq}{\end{equation}}

\input{epsf}

\begin{document}

\tolerance 10000

\newcommand{\vk}{{\bf k}}
\newcommand{\nd}{^{\vphantom\dagger}}
\parskip=0pt

\title{Theory of the Three Dimensional Quantum Hall Effect in Graphite}

\author{B. Andrei Bernevig$^1$, Taylor L. Hughes$^2$, Srinivas Raghu$^2$ and Daniel P. Arovas$^{2,3}$}

\affiliation{$^1$Princeton Center for Theoretical Physics, Princeton
University, Princeton, NJ 08544} \affiliation{$^2$Department of
Physics, Stanford University, Stanford, CA 94305}
\affiliation{$^3$Department of Physics, University of California at
San Diego, La Jolla, CA 92093}

\begin{abstract}
We predict the existence of a three dimensional quantum Hall effect
plateau in a graphite crystal subject to a magnetic field. The
plateau has a Hall conductivity quantized at $\frac{4e^2}{\hbar}
\frac{1}{c_0} $ with $c_0$ the $c$-axis lattice constant. We analyze
the three-dimensional Hofstadter problem of a realistic
tight-binding Hamiltonian for graphite, find the gaps in the
spectrum, and estimate the critical value of the magnetic field
above which the Hall plateau appears. When the Fermi level is in the
bulk Landau gap, Hall transport occurs through the appearance of
chiral surface states.  We estimate the magnetic field necessary for
the appearance of the three dimensional quantum Hall Effect to be
$15.4\,$T for electron carriers and $7.0\,$T for hole carriers.
\end{abstract}

\date{\today}

\pacs{72.25.-b, 72.10.-d, 72.15. Gd}

\maketitle

Recent advances in the fabrication of single graphene sheets as well
as the striking initial experiments on the relativistic quantum
Hall effect in graphene \cite{novoselov2005,zhang2005} have
generated intense interest in this remarkable material. Most
of the theoretical and experimental research has focused on the
properties of the low energy excitations close to half filling which
have a Dirac spectrum with a speed of light of the order of $10^6\,
{\rm m}/{\rm s}$. The spin-unpolarized quantum Hall effect shows a
sequence of plateaus at $\sigma_{xy} = (n+\frac{1}{2})\times
4\frac{e^2}{h}$ consistent with the existence of two Dirac cones as
well as a spin degeneracy \cite{novoselov2005,zhang2005}.

The phenomenon of Hall conductivity quantization is however not
restricted to two dimensions and can occur in bulk samples, albeit
under more stringent conditions. It was first observed by Halperin
\cite{halperin1987} that for a three-dimensional (3D) electron system
in a periodic potential, if the Fermi level lies inside an energy gap, the
conductivity tensor is necessarily of the form:
\begin{equation}
\sigma\nd_{ij} =\frac{e^2}{2 \pi h}\, \epsilon\nd_{ijk} G\nd_k
\end{equation}
\noindent where $\epsilon_{ijk}$ is the fully antisymmetric tensor
and $\vec{G}$ is a reciprocal lattice vector of the potential which
may or may not be zero. The periodic potential can be
generated by either an underlying lattice or spontaneously by many-body
effects through the formation of density waves.

The existence of the three dimensional quantum Hall effect (3DQHE)
has not, however, been observed in any bulk materials, although the
effect has been observed in stacked superlattices of 2D
semiconductor layers, each exhibiting a 2DQHE, where the inter-layer
tunneling is engineered away \cite{stormer1986}. The only current
bulk material candidates for 3DQHE are the Bechgaard salts
\cite{balicas1995,mckernan1995}. The difficulty inherent in
realizing the 3DQHE in a specific bulk material can be understood by
considering the spectrum of an electron in a periodic potential
under an applied magnetic field. In 2D, this Hofstadter problem
\cite{hofstdater1976} results in magnetic energy bands.  Associated
with each band is an integer quantized Hall conductance; when
the Fermi level lies in a gap, the observed Hall conductance is a
sum over the contributions from each filled band.  In 3D, the
momentum component parallel to the magnetic field disperses freely
and the very existence of such gaps is not guaranteed; indeed it is
exceptional. Tilting the magnetic field away from the
crystallographic axes helps \cite{koshino2002}, but does not
guarantee, the opening of gaps.

\begin{figure}
\includegraphics[width=2.8in]{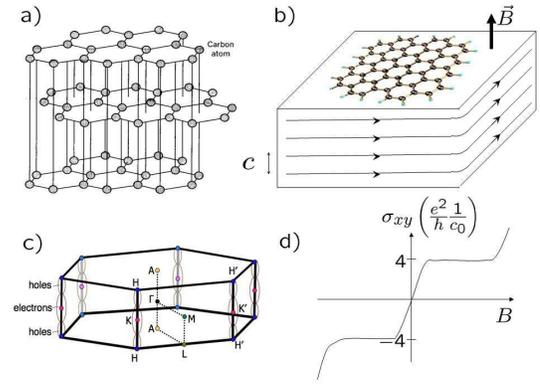}
\caption{(a) Graphite in Bernal stacking. (b) Under strong magnetic
field, graphite is gapped in the bulk and exhibits chiral surface
sheet states. (c) Idealized Brillouin zone for graphite. (d)
Predicted 3D Hall conductivity, quantized in units of $1/c_0$. Only
one plateau is observable in graphite.}\label{graphite}
\end{figure}

In this paper we show that a true, bulk 3DQHE is realized in
graphite under a large magnetic field parallel to the $c$-axis.
Three factors conspire to render this possible: the large Landau gap
of the integer quantum Hall state in graphene, the weak interplane
hopping, and the Bernal stacking.  We first give a physical argument
for the existence of 3DQHE based on adiabatic continuity, then
perform a full Hofstadter calculation in 3D of the band and surface
states structure, using the realistic Johnson-Dresselhaus
\cite{johnson1975,leung81} Hamiltonian for graphite, plus a magnetic
field.  We find the minimum magnetic field necessary for a 3DQHE to
be $15.4\,$T for electrons and $7.0\,$T for holes, and show that
only \emph{one} Hall plateau (Fig. \ref{graphite}) will be observed
due to band closing in the higher Landau levels (LLs). Besides the
obvious prediction of a plateau in off-diagonal conductive response,
we also predict that a correlated chiral surface state
\cite{balents1996}, occurs at the boundary of the sample.  Based on
the recent experimental focus on graphite \cite{zhou2006} we believe
our prediction is testable with current experimental techniques.
Previous Hall plateaus observed in graphite
\cite{kempa2006,kopelevich2003} are either even in $B$, or come in
multiple sequences consistent with the graphene QHE plateaus
\cite{novoselov2004,kopelevich2003}, and are hence different than
our prediction of a \emph{single} quantized Hall plateau of Hall
conductivity twice as large as the one observed in graphene for
Fermi level in the first Landau gap.

\begin{figure}
\includegraphics[width=3.5in]{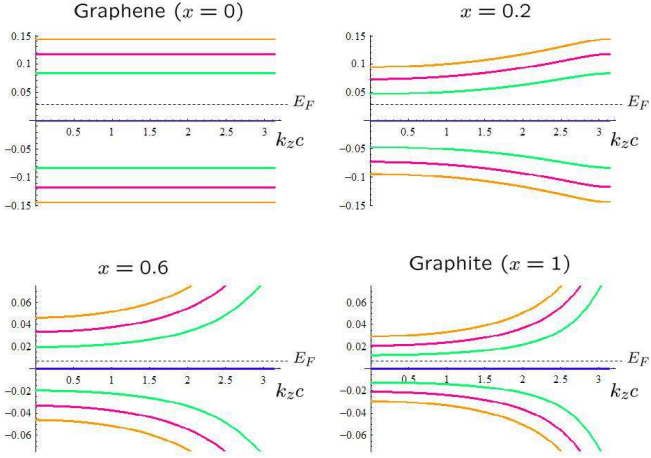}
\caption{ Spectrum of the zero mode and first few Landau levels of
our toy model in $B=10\,$T interpolated between graphene ($x=0$), where they
do not $k_z$ disperse, and graphite ($x=1$). The zero mode (blue) is doubly
degenerate, giving a 3D Hall conductance of $\frac{e^2}{h}
\frac{1}{c_0}$ per independent $K$ point per spin.
}\label{toymodelLL}
\end{figure}

Graphite is a layered material consisting of weakly coupled parallel
graphene layers in an $ABABAB$ configuration known as Bernal stacking
\cite{dresselhaus2002}, as depicted in Fig. \ref{graphite}.  Let us
initially set the inter-layer hopping to zero and place a large
number of uncoupled stacked graphene layers in a strong
magnetic field perpendicular to their surface. Each layer exhibits a
relativistic quantum Hall Effect as previously described with a
LL energy $E_n = \pm\sqrt{2n}\,\hbar v/\ell$, where $\ell=\sqrt{\hbar c/eB}$
is the magnetic length, $n=0,1,\dots$. and $v \approx 10^6\, {\rm m}/{\rm s}$.
Typical values of the gap are roughly $0.1$ eV for $B=10\,$T and $>0.25$ eV
for $B>40\,$T, thus making graphene the first system to exhibit
quantized Hall conductance at room-temperature.  If we place the
Fermi level in the first Landau gap, our system of uncoupled
graphene layers trivially exhibits a 3DQHE, with the bulk enveloped
by a sheath of chiral surface states as in Fig. \ref{graphite}.
For uncoupled layers, there is no $k_z$ dispersion.  We now adiabatically
turn on intra-layer hopping, which causes almost all LLs to
disperse with $k_z$. The exception are the zero
modes which are stable in an idealized model due to particle-hole
symmetry, as we show below. By adiabatic continuity, the sheath of
chiral states (and hence the 3DQHE) must be stable as long as the
Landau gap does not collapse.

We first show that the statement above is true for a simplified
toy-model of graphite. If we introduce a parameter $x$ which
interpolates between uncoupled graphene layers ($x=0$) and graphite
($x=1$), the minimal Hamiltonian of our system is a $4\times 4$
Hermitian matrix acting on the lattice spinor $(\psi\nd_{\rm A},
\psi\nd_{\rm B},\psi\nd_{\rm C}, \psi\nd_{\rm D})$, where (A,B) and
(C,D) are the two inequivalent sites of the lower and upper graphene
layers in the Bernal stacking, respectively.  Its nonzero
independent elements are ${\cal H}\nd_{\rm AB}=-{\cal H}^*_{\rm
CD}={\sqrt{3}\over 2}\, t\nd_\parallel\,a\nd_0\,k\nd_-$, and ${\cal
H}\nd_{\rm AC}=-2x t\nd_\perp \cos({1\over 2}k_z c\nd_0)$. The
in-plane hopping is $t_\parallel = 3.16$ eV, $t_\perp = 0.39$ eV is
the interlayer (A-C) hopping in graphite ($x=1$)
\cite{dresselhaus2002}, and $a\nd_0= 2.456$ \AA\ is the graphene
lattice constant (in-plane A-A distance). The $c$-axis lattice
constant is $c\nd_0= 6.74\,$\AA\ (the distance between consecutive
lattice planes equal to ${1\over 2}c\nd_0$), and $k_\pm=k_x\pm
ik_y$. The in-plane dispersion is expanded about the point $K =\big(
\frac{ 4\pi}{3 a_0},0,0 \big)$, which is the Dirac point in
graphene. In a uniform magnetic field, adopting the symmetric gauge
$\vec{A}= \frac{1}{2}B (- y,x,0)$, the Kohn-Luttinger substitution
$k_\pm \to k_x\pm i k_y + {e\over\hbar c}(A_x\pm iA_y)$
\cite{mcclure1960}. In the specific case of ($x=1$) this model is
the same as the one used in \cite{nilsson2006}. Let $b^\dagger, b$
be creation and annihilation Landau Level operators (with
$[\,b,b^\dagger] =1$), and introduce the notation
$c\nd_z=\cos({1\over 2}k\nd_z c\nd_0)$. The Hamiltonian is
\begin{equation}
{\cal{H}}(x) = \left(%
\begin{array}{cccc}
  0 & -\epsilon\,t\nd_\parallel\, b & -2x\,t\nd_\perp c_z & 0 \\
-\epsilon\,t\nd_\parallel\, b^\dagger & 0 & 0 & 0 \\
  -2x\, t\nd_\perp c_z & 0 & 0 & -\epsilon\,t\nd_\parallel\,  b^\dagger \\
  0 & 0 & -\epsilon\,t\nd_\parallel\, b & 0 \\
\end{array}\right) \label{toyhamilt}
\end{equation}
\noindent with $\epsilon=B/B\nd_0$ and $B\nd_0=(hc/e)/3\pi a_0^2=7275\,$T,
and is diagonalized in a basis $\psi =(|n \rangle, |n+1
\rangle, |n \rangle, |n-1 \rangle)$, for $n>0$.  We obtain below the $4$ LL
energies $E_n$ as
\begin{align}
E_n=&\pm\bigg[ (n+\textstyle{{1\over 2}})\, \epsilon^2\, t_\parallel^2+2x^2t_\perp^2 c_z^2\\
&\quad\pm\sqrt{{\textstyle{1\over 4}\, \epsilon^4\, t_\parallel^4}+ 4\,(n+\textstyle{{1\over 2}}) \,x^2\,\epsilon^2\, t_\parallel^2\,
t_\perp^2\, c_z^2+ 4x^4\, t_\perp^4\, c_z^4} \bigg]^{1/2}\ ,\nonumber
\end{align}
This spectrum has explicit particle-hole symmetry. For $n=0$ there
are eigenvalues at
$\pm\big(\epsilon^2\,t_\parallel^2 + 4 x^2\,t_\perp^2\,c_x^2\big)^{1/2}$
and a doubly degenerate level at $E\nd_0=0$.  (All levels receive an
additional double degeneracy owing to the existence of the
inequivalent $K$ point.)  The next Landau bands are the lowest two
energy levels of $n=1$.  Using this spectrum we find that the gap between
the zero mode and the LL above or below it cannot collapse
upon interpolating between $x=0$ and $x=1$ for \emph{any} value of
the magnetic field. As such, by adiabatic continuity, the Hall
conductance when the Fermi level is in the 3D gap (with doubling
for the two spins) is:
\begin{align}
\sigma_{xy} &= \frac{4e^2}{h}\! \int\!\! {d^3\! k\over (2 \pi)^3}\,
{\rm Im}\>\Big\langle{\partial\psi\over\partial k_x} \Big|{\partial\psi\over\partial k_y} \Big\rangle = \nonumber \\
&=(2n+1)\,\frac{4e^2}{h} \int\frac{d k_z}{2 \pi} = (2n+1) \times\frac{4e^2}{hc_0}\ .
\end{align}
\noindent  For the Fermi level between the zero mode and the first
LL, $n=0$ and $\sigma_{xy} = \pm  4e^2/hc_0$.  The Bernal
stacking of graphite accounts for the extra factor of $2$ relative
to graphene and $\int\!\frac{d^2 k}{(2 \pi)^2}\, {\rm Im}\,
\langle\partial\nd_{k_x} \psi\, |\,\partial\nd_{k_y} \psi \rangle =
2n+1$ is the TKNN integer \cite{TKNN82} of the relativistic graphene
bands when the Fermi level is placed in the $n^{\rm th}$ bulk gap.
We observe that the gap between the first and the second LL in
graphite closes for any realistic value of the $B$ field (Fig.
\ref{toymodelLL}), and hence higher $n$ plateaus will not be
observed.  Zeeman splitting could give rise to a $\sigma_{xy}= 0$
plateau, but is smeared by the dispersion of the zero mode in the
realistic graphite model used below, and hence the predicted Hall
conductance is sketched in Fig. \ref{graphite}.

The existence of a full gap in the 3D LL spectrum is an artifact of
the toy model involving only $t\nd_\parallel$ and $t\nd_\perp$.  It
derives from the presence of a flat, twofold degenerate band at the
Fermi level along the HK spines of the Brillouin zone.  The Bernal
stacking is crucial, for if the individual graphene planes were
stacked identically, all LLs would acquire a $k_z$-dependent contribution
to their energies $-2t_\perp \cos(k_zc\nd_0)$, and would then overlap
at all but extremely high magnetic
fields \cite{rhombo}.  More realistic models for graphite, such as the
Slonczewski-Weiss-McClure (SWMC) model
\cite{slonczewski1958,mcclure1957} or that of Johnson and
Dresselhaus (JD) \cite{johnson1975,leung81}, contain small $c$-axis B-B
(D-D) hopping terms, through the open hexagons of the CD (AB) plane.
Their value is small on the scale of nearest neighbor hopping --
only $10$ meV, leading to a bandwidth of $40$ meV along
the HK spine.  But the presence of such terms is crucial toward
understanding the properties of graphite.  At $B=0$, they result in
semimetallic behavior, whereas the toy model incorrectly predicts a
zero gap semiconductor.  For weak fields, they lead to overlap of
the LLs and destruction of the QHE. However, as we shall
show, the principal gaps surrounding the central $n=0$ LLs
survive for $B>7.0\,$T (holes) and $B>15.4\,$T (particles).  Lightly
doped graphite, then, will exhibit a 3DQHE at these fields.  We next
describe our solution of the JD model
in a magnetic field on a torus and a Laughlin cylinder,
finding the LLs and the surface states, and determining
the critical fields $B_{\rm c}$ at which energy gaps open across the
entire Brillouin zone \cite{foot}.

\renewcommand\arraystretch{1.8}
\begin{table}
\caption{Tight-binding parameters}
\label{JDM}
\smallskip
\begin{tabular}{||c|r||c|r||c|r||}\hline\hline
Parameter & meV & Parameter & meV & Parameter & meV \\ \hline\hline
$t^{\parallel,1}_{\rm AB}=t^{\parallel,1}_{\rm CD}$ & $4200$ & $t^{\parallel,2}_{\rm AB}=t^{\parallel,2}_{\rm CD}$ & $512.5$ &
$t^{\parallel,3}_{\rm AB}=t^{\parallel,3}_{\rm CD}$ & $15$ \\ \hline
$t\nd_{\rm AC}$ & $-390$ & $t\nd_{\rm BD}$ & $-315$ & $t\nd_{\rm AD}=t\nd_{\rm BC}$ & $-44$  \\ \hline
$t^\perp_{\rm AA'}=t^\perp_{\rm CC'}$ & $-19$ & $t^\perp_{\rm BB'}=t^\perp_{\rm DD'}$ & $10$ & $\Delta$ & $50$\\
\hline\hline
\end{tabular}
\end{table}

\begin{figure}
\includegraphics[width=3.3in]{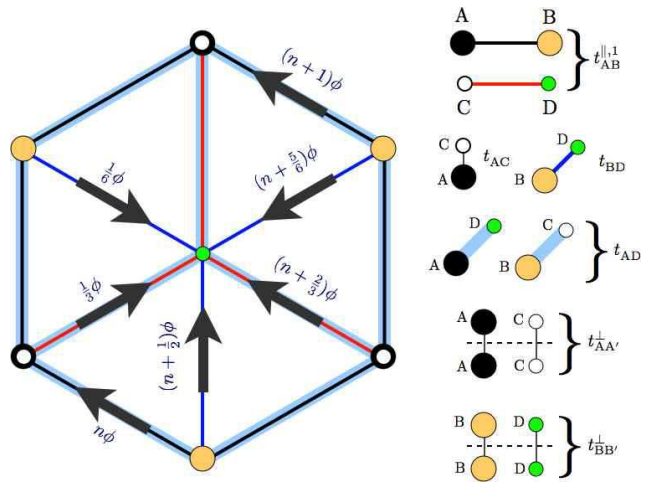}
\caption{Definition of JD hopping matrix elements and flux
assignment.  Sites $A$ and $B$ belong to different graphene layers
than sites $C$ and $D$, and Bernal stacking corresponds to sites
$A$ and $C$ differing by a $c$-axis translation.  Not shown
are the further-neighbor in-plane hoppings $t^{\parallel,2}_{\rm AB}$
and $t^{\parallel,3}_{\rm AB}$.  $\phi=2\pi p/q$ is the magnetic flux per
hexagon in units of $\hbar c/e$, and $n$ runs from $1$ to $q$, the number
of hexagons in a magnetic unit cell.}\label{gauge}
\end{figure}

The JD model \cite{johnson1975,leung81} is a tight-binding Hamiltonian derived
from the ${\bf k}\cdot{\bf p}$
theory of SWMC.  Its nine parameters are given in Tab. \ref{JDM}.  In addition to
nearest neighbor hoppings, there are also further neighbor hoppings, both in-plane
(extending to third and fourth neighbor) and between planes.  There is also an
energy asymmetry $\Delta=\varepsilon\nd_{\rm A(C)}-\varepsilon\nd_{\rm B(D)}$.
We then introduce a magnetic field through Peierls phases along the links, consistent
with all rotational (and screw axis) symmetries of the underlying lattice, as depicted
in Fig. \ref{gauge}.

We solve the model through a combination of exact diagonalization,
Lanczos method (for $q>1000$, where the flux per hexagonal plaquette
is $1/q$ Dirac quanta $\phi\nd_0=hc/e$), and low field expansion.  For the bulk band
structure, we impose doubly periodic ({\it i.e.\/} toroidal)
boundary conditions in the $(x,y)$ plane, while to study edge
(surface) states we impose singly periodic ({\it i.e.\/} cylindrical)
boundary conditions. The latter case may also be solved via a
transfer matrix formalism \cite{hatsugai1993}.  Due to the algebraic
complexity of the problem, we postpone the explicit details of the
$B$-field Hamiltonian and transfer matrix for a future publication.
For $q>20$ the Hofstadter broadening becomes negligible and the band
energies as a function of $k_x$ become non-dispersive, corresponding
to the real situation in which the magnetic field splitting is small
compared to the \emph{in-plane} hopping amplitude.

\begin{figure}
\includegraphics[width=3.0in]{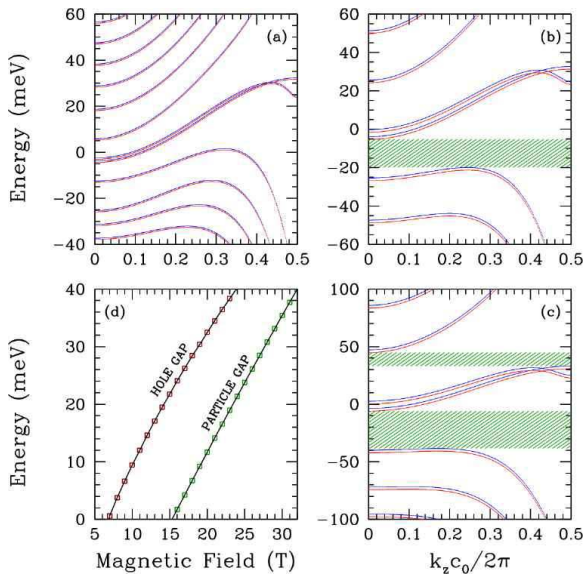}
\caption{Clockwise from upper left: (a) $B=5\,$T, no gap present in the
full spectrum. (b) $B=12\,$T, clear gap in the hole LL spectrum.
(c) $B=20\,$T clear gaps for both hole and electron LL.  Spin-up (blue)
and spin-down (red) bands are shown.  (d) Principal energy gaps surrounding
$n=0$ LLs, including effects due to Zeeman splitting.  The
particle gap collapses at $B_{\rm c}^{\rm e}=15.4\,$T and the hole
gap at $B_{\rm c}^{\rm h}=7.0$\,T.  All energies have been shifted
upward by $100\,$meV.}
\label{gaps}
\end{figure}

\begin{figure}
\includegraphics[width=3.5in]{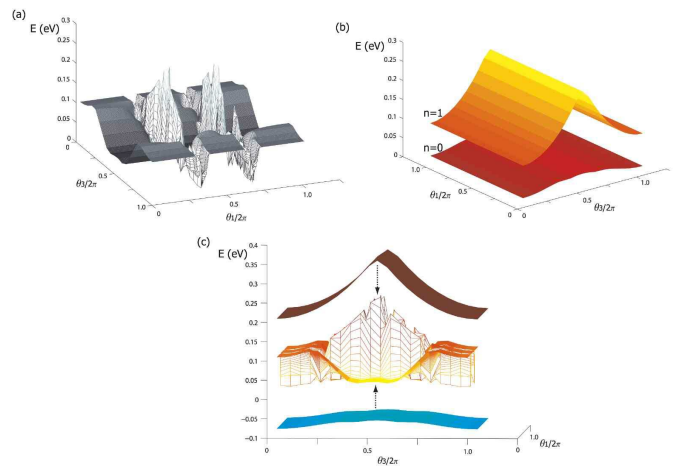}
\caption{Zero mode, first electron Landau level and surface states
spectrum over the full Brillouin zone for  one spin species at $B=40\,$T.
(a) Surface states spectrum (b) Zero mode and first electron Landau
level spectrum (c) Surface states and bulk Landau levels together.
The $n=1$ ($n=0$) levels are shifted by $+0.1\,$eV ($-0.1\,$eV) for clarity.
There are two surface states per spin species crossing the bulk gap.}
\label{surfacestates}
\end{figure}

We observe the following characteristics of the spectrum. The zero
mode has a dispersion of $40\,$meV in the low field limit. The first
LL, however, disperses strongly as a function of $k_z$ and
for each value of $q$ we scan the energy $E(k_z)$ spectrum to look
for the smallest gap.  Fig. \ref{gaps} shows the bulk Landau Level
spectrum for $B=5$, $12$, and $20$ Tesla, as well as the field-dependence of
the principal gaps surrounding the central, weakly dispersive $n=0$
LLs. We find that both gaps are indirect for $B<30\,$T. In
Fig. \ref{surfacestates} we plot the bulk band and surface state
spectrum as a function of $k_x$ and $k_z$, which proves the
existence of gaps and surface states over the entire Brillouin zone.
On the Laughlin cylinder, for each value of $k_z$, we obtain $2$
edge states on each of the upper/lower edge of the cylinder. Unlike
the low-field bulk LLs, the edge states disperse as a
function of $k_x$ and cross the bulk gap to give $2e^2/hc\nd_0$
per spin. We have checked that our numerical results match the
theoretical analysis of the continuum $k$ approximation of the JD
model.

At low fields, each LL accommodates
$\frac{\sqrt{3}}{2}a_0^2B/4\phi\nd_0= 3.16\times 10^{-6}\,B\,[{\rm T}]$
states per carbon atom.  Accounting for the quadruple degeneracy of
the LLs (two $K$ points and two spin polarizations), the
central $n=0$ levels will be filled for fields below $20\,$T  at a
doping of only $0.025\%$.  For the lowest field for which we
predict the effect, the doping is a modest $0.01 \%$.  Unlike in a
many-body gap, the Fermi level is not pinned and the width of the
LL will be given by the width of the mobility
single-particle gap in disordered graphite.  Strong disorder leads
to wide Hall plateaus, and weak disorder to narrow
ones \cite{chalker1995}.

Besides the prediction of a quantized response, the chiral surface
sheet should exhibit ballistic in-plane longitudinal response
$\rho_{xx} \rightarrow 0$ as $T \rightarrow 0$.  Remarkable transport
properties are also present along the direction parallel to the
magnetic field \cite{balents1996}. In this direction the system is a
stable metal with a temperature independent resistivity $\rho_{zz}$,
with a value which can be much larger than $h/e^2$, although,
unlike previously predicted \cite{balents1996} we do expect the
metallic phase to be unstable to
very strong disorder and impurity concentration which levitate the
bulk extended states above the Fermi level \cite{laughlin1984}.

In conclusion, we predict the appearance of \emph{one} plateau of 3D
quantized Hall transport in doped bulk graphite under strong
magnetic fields parallel to its $c$-axis.  We analyzed the
Hofstadter problem in graphite and estimate the minimum field to be
approximately $7.0\,$T (for holes) and $15.4\,$T (for electrons).
While our general conclusions are robust, differences among material
parameters reported in the literature can shift the critical fields by
several Tesla.  The recent advances and interest in graphene and graphite
make our prediction testable with current experimental techniques.

B. A. B. wishes to thank F.D.M. Haldane for numerous challenging
discussions.  D. P. A. is grateful to the Condensed Matter Theory
group at Stanford University for its generous hospitality and
sabbatical support, and to S.-C. Zhang for discussions.


\begin{thebibliography}{26}
\bibitem[{rho()}]{rhombo}
\bibinfo{howpublished}{We have found that no 3DQHE exists in rhombohedral
  graphite, with $ABCABC$ stacking (details to be presented in a future
  publication).}
\end{thebibliography}
\end{document}